\documentclass{pnastwo}

\usepackage{graphicx}
\usepackage{amssymb}
\usepackage{amsmath}
\usepackage{amsfonts}
\usepackage{mathrsfs}

\begin{document}

\title{Generalized Khinchin Theorem for a Class of Aging Processes}

\author{S. Burov%
\affil{1}{Department of Physics, Bar Ilan University, Ramat-Gan 52900,
Israel},
R. Metzler%
\affil{2}{Physics Department, Technical University of Munich, D-85747
Garching, Germany},
E. Barkai%
\affil{1}{Department of Physics, Bar Ilan University, Ramat-Gan 52900,
Israel}}

\maketitle


\begin{article}

\begin{abstract}
The Khinchin theorem provides the condition that a stationary process is
ergodic, in terms of the behavior of the corresponding correlation function.
Many physical systems are governed by non-stationary processes in which
correlation functions exhibit aging. We classify the ergodic behavior of such
systems and provide a generalization of Khinchin's theorem. Our work quantifies
deviations from ergodicity in terms of aging correlation functions.
Using the framework of the fractional Fokker-Planck
equation we obtain a simple analytical expression for the two-time correlation
function of the particle displacement in a general binding potential, revealing
universality in the sense that the binding potential only enters into the
prefactor through the first two moments of the corresponding Boltzmann
distribution. We discuss applications to experimental data from systems
exhibiting anomalous dynamics.
\end{abstract}


\dropcap{T}he dynamics of physical observables is commonly quantified in
terms of time dependent correlation functions; these are pair products of
dynamic observables $O(t_1)$ and $O(t_2)$, that are averaged over an ensemble,
$C(t_2,t_1)=\langle O(t_2)O(t_1)\rangle$. When such a correlation function
describes a stationary process, $C(t_2,t_1)$ is a function of the time
difference only, that is, $C(t_2,t_1)=C(|t_2-t_1|)$. For such processes the
correlation function contains the dynamical information on other physical
observables via fundamental theorems \cite{Forester}. For instance, the
Wiener-Khinchin theorem relates the power spectrum to the correlation
function; or the fluctuation-dissipation theorem connects correlation
functions to linear response functions. Another well known
example is Khinchin's theorem \cite{Khinchin}, that provides a criterion
for ergodicity of a process in terms of the corresponding stationary
correlation function. However, the stationarity property
is found to be violated in numerous systems (see below). These systems exhibit
aging properties, that are intimately connected to the non-stationary behavior.
Correlation functions in aging systems behave very differently from their
stationary counterparts \cite{Glasses,Grigolini01,barsegov,witko}. For
instance, such
non-stationary systems may be characterized by correlation functions of the
type $C(t_2,t_1)=h(t_1/t_2)$ (if $t_2\geq t_1$), i.e., the two times enter
as a quotient rather than their difference. This non-stationary behavior
is also connected to  a breaking of ergodicity in the sense that long time
averages differ from ensemble averages of the same quantities
\cite{Glasses,Bouchaud,bouchaud1,odagaki,Bel}.
The relation between correlation functions and
ergodicity breaking can be quantified by the Edwards-Anderson parameter
\cite{Glasses}, see below.

We here present a generalization of the Khinchin theorem for aging systems,
relating fluctuations of time averages to the corresponding aging correlation
function, that we calculate for a class of important stochastic processes. In
particular we derive an analytical expression for the two-time position
correlation function in the presence of an external binding potential $U(x)$
based on the the general framework of the fractional Fokker-Planck equation
\cite{Metzler,MetzlerKlafter,report}. The latter describes systems in which
the mean squared displacement in free space scales as $\langle x^2(t)\rangle
\propto t^{\alpha}$, where the anomalous diffusion exponent $\alpha$ ranges in
the interval $0<\alpha<1$ \cite{MetzlerKlafter,report,Bouchaud01}.
Our results for the correlation function
are quite general and could also be applied to continuous time random walk
(CTRW) systems in confined geometries and mean field descriptions of models
such as the quenched trap model \cite{Bouchaud01}. In particular, we show that
for sufficiently long times the \emph{correlation function behaves universally},
and the dependence on the potential $U(x)$ enters only in the prefactor through
the first two moments of the corresponding Boltzmann distribution.
We also discuss aging properties of the correlation function and
the time averaged mean squared displacement.
Moreover we demonstrate agreement of our results with experimental data.

Physical systems displaying non-stationary behavior like aging and ergodicity
breaking traditionally included glassy systems such as spin glasses
\cite{Glasses}, colloidal glasses \cite{mattson}, gels \cite{Weitz},
turbulent systems \cite{Grigolini02}, or tracer dispersion in subsurface
hydrology \cite{scher}, among others. More recently advanced single-molecule
experiments reveal other types of complex systems with similar behavior.
These are systems exhibiting anomalous diffusion and slow, non-exponential
relaxation dynamics \cite{Bouchaud01,MetzlerKlafter,report,irwin}. For
instance, they include blinking quantum dots \cite{Dahan,pt,margo,margo1}, or
biologically relevant systems. The latter include subdiffusion of tracer
particles in living biological cells \cite{Golding,b1,b2,b3,b4,b5} or
reconstituted crowding systems \cite{reconst,r1,r2,r3}, protein conformational
dynamics \cite{proteins}, or the motion of bacteria in a biofilm \cite{biofilm}.
Here we show how the knowledge of the aging correlation function allows us to
quantify the non-ergodic behavior of the process.

\subsection{Correlation functions and ergodicity}

The concept of ergodicity plays a central role in statistical physics.
Boltzmann's ergodic hypothesis relates the time average of a physical
observable $O$ to its ensemble average and states that both are equal
in the long time limit:
\begin{equation}
\overline{O}\equiv\lim_{t\to\infty}\frac{1}{t}\int_0^tO(t')dt'=\langle
O\rangle.
\label{ergod1}
\end{equation}
where $\langle\cdot\rangle$ represent an ensemble average with respect
to a steady state probability distribution.
The connection between the ergodic hypothesis and correlation functions was
established by Khinchin \cite{Khinchin}, a work of considerable impact to
statistical physics \cite{Kubo}. Khinchin's theorem states that an observable
$O(t)$ is ergodic if the associated correlation function is ``irreversible'',
in the sense that if $O(t)$ fulfills
\begin{equation}
\label{ergod2}
\lim_{\Delta\to\infty}\langle O(t_1+\Delta)O(t_1)\rangle=\langle O\rangle^2
\qquad \left(\Delta=|t_2-t_1|\right),
\end{equation}
then Eq.~\eqref{ergod1} holds. In the derivation of Khinchin's theorem it is
assumed that the process is stationary and the system reached a steady state,
$C(t_2,t_1)=\langle O(t_2)O(t_1)\rangle=C(\Delta)$. Thus the irreversibility
condition corresponds to the requirement $C(\Delta)=\langle O\rangle^2$ for
$\Delta\to\infty$. Note that for stationary systems \cite{Lee} irreversibility
is a broader concept than the ergodic hypothesis, and that therefore
Khinchin's theorem cannot always be reversed. The case of power-law decaying
correlation functions was treated previously based on
the generalized Langevin equation \cite{Hanggi,Oliviera}. This gives rise to
a stationary process, such that Khinchin's theorem holds for this kind of
anomalous dynamics.

Having in mind the aforementioned non-stationary systems, the following
question arises: Does irreversibility imply ergodicity also for aging systems?
And if not, what theorem replaces Khinchin's? To answer these questions we
quantify ergodicity in terms of second moments, namely if
$\left<\overline{O}^2\right>=\langle O\rangle^2$ the dynamics is ergodic, as
originally pointed out by Khinchin and $\left<\overline{O}\right>=\langle
O\rangle$ always holds for processes where $\langle
O(t)\rangle$ is constant for long enough time. From Eq.~\eqref{ergod1}
we find for the second moment
\begin{equation}
\left<\overline{O(t)}^2\right>=\frac{1}{t^2}\int_0^t\int_0^t\langle O(t_2)
O(t_1)\rangle\,dt_2\,dt_1.
\label{ergod3}
\end{equation}
Assuming aging behavior for the correlation function in the above form
$\langle O(t_2)O(t_1)\rangle=h(t_1/t_2)$ ($t_2\geq t_1$) (i.e., we use
the aging regime as the ``steady state'' of the system similarly to invoking
stationarity in Khinchin's theorem) Eq.~\eqref{ergod3} becomes
\begin{equation}
\left<\overline{O(t)}^2\right>=\frac{2}{t^2}\int_0^t\int_0^{t_2}h(t_1/t_2)
dt_1dt_2.
\label{ergod4}
\end{equation}
After the substitution $t_1^2+t_2^2=r^2$ and $t_1/t_2=\tan(\theta)$ we find
\begin{equation}
\left<\overline{O}^2\right>=\int_0^1h(z)\,dz.
\label{ergod5}
\end{equation}
This is a general relation between the fluctuation of the time averaged
and ensemble averaged correlation functions. For ergodicity to hold, we
consequently require that in the long time limit the condition
 \begin{equation}
\left<O\right>^2=\int_0^1h(z)\,dz
\label{ergod6}
\end{equation}
is fulfilled. Simultaneously we can rewrite the irreversibility condition
$\lim_{\Delta \to \infty}h(t\big/t+\Delta)=\left<O\right>^2$ in the form
\begin{equation}
\lim_{z\to 0}h(z)= \left<O\right>^2.
\label{ergod7}
\end{equation}
Thus, to fulfill Eq.~\eqref{ergod6} the condition \eqref{ergod7} is not
sufficient, and Khinchin's theorem does not hold for aging systems. Namely
irreversibility does not imply ergodicity in aging system. We
define a generalization of Khinchin's theorem stating that in the case of
irreversible dynamics the condition for ergodicity in an aging system is
given by 
\begin{equation}
\lim_{z\to 0}h(z) = \int_0^1h(z)\,dz.
 \label{ergod_thoer}
\end{equation}
The knowledge of the correlation function resolves the question of ergodicity
for a system. More importantly, by help of Eq.~\eqref{ergod5} the correlation
function allows one to quantify the fluctuations of the time averages and
ergodicity breaking:
\begin{equation}
\left<\overline{O}^2\right>-\left<O\right>^2=\int_0^1\left[
h(z)-h(0)\right]\,dz,
\label{ergod8}
\end{equation}
where we assumed irreversibility. We now turn to the calculation of the
correlation function in an aging system and prove that it is irreversible.
The result will be shown to exhibit universal features which, together with
Eq.~\eqref{ergod8}, imply a generic behavior of the fluctuations of time
averages.

\section{Model}

\subsection{Fractional dynamics}

Anomalous diffusion in an external potential $U(x)=-\int^xF(x')dx'$
is governed by the fractional Fokker-Planck equation
\cite{Metzler,MetzlerKlafter}
\begin{equation}
\frac{\partial f(x,t)}{\partial t}=K_{\alpha}\,_0D_t^{1-\alpha}\mathbb{L}_{
\mathrm{FP}}f(x,t),
\label{ffpe01}
\end{equation}
in which the Fokker-Planck operator
$\mathbb{L}_{\mathrm{FP}}=-\frac{\partial}{\partial
x}\left[\frac{F(x)}{k_BT_{mp}}
\right]+\frac{\partial^2}{\partial x^2}$
includes drift and diffusion terms \cite{Risken}, $K_{\alpha}$ is the anomalous
diffusion coefficient of dimension $\mathrm{m}^2/\mathrm{sec}^{\alpha}$, and
$k_BT_{mp}$ the thermal energy. The fractional Riemann-Liouville operator
\begin{equation}
_0D_t^{1-\alpha}z(t)=\frac{\partial}{\partial t}\frac{1}{\Gamma(\alpha)}
\int_0^t\frac{z(t')}{(t-t')^{1-\alpha}}dt',\,\,\,(0<\alpha<1)
\end{equation}
involves long-range memory effects \cite{MetzlerKlafter,report}.
Eq.~\eqref{ffpe01} describes the time evolution of the
single-time probability density function $f(x,t)$ and has been studied
extensively for different potentials \cite{MetzlerKlafter,report,Barkai02}.
Eq.~\eqref{ffpe01} can be derived as the long-time limit of a continuous
time random walk model, in which the local probability to jump left and
right is biased by the external potential $U(x)$ \cite{ffpederive,elis}.

The fractional Fokker-Planck equation \eqref{ffpe01} can be rephrased in
terms of the Langevin equations \cite{Fogedby,Friedrich00,Weron,Weron1}
\begin{equation}
\frac{dx(s)}{ds}=\frac{K}{k_B T}F(x)+\eta(s)\quad\mbox{(a)},\quad
\frac{dt(s)}{ds}=\omega(s)\quad\mbox{(b)},
\label{subordination1}
\end{equation}
where $s$ is an internal time (unit-less) and $t$ is the physical (laboratory)
time. In Eq.~\eqref{subordination1} the noise $\eta(s)$ is white and Gaussian
with zero mean $\langle\eta(s)\rangle=0$ and autocorrelation $\langle\eta(s)
\eta(s')\rangle=2K\delta(s-s')$; $K>0$ is the diffusivity for the normal
diffusion process in internal time $s$.
Conversely the noise $\omega(s)$ represents an asymmetric L\'{e}vy-stable
process of order $\alpha$ such that the probability density function of $s$
is \cite{Barkai02,Friedrich00,Weron}
\begin{equation}
p_t(s)=\frac{1}{\alpha}\left(\frac{K_\alpha}{K } \right)^{1/\alpha}
\frac{t}{s^{1+1/\alpha}}
L_{\alpha}\left(\left(\frac{K_\alpha}{K }
\right)^{1/\alpha} \frac{t}{s^{1/\alpha}} \right).
\label{pts}
\end{equation}
$L_{\alpha}$ is a one-sided L\'{e}vy stable probability density function
with Laplace transform $\exp(-\lambda^\alpha)$ \cite{Barkai02}. 
The representation of the fractional dynamics by means of the coupled Langevin
equations \eqref{subordination1} is usually termed subordination
\cite{Fogedby}. 
With Eq.~(\ref{pts}) it is easy to show~\cite{Barkai02} that statistical
properties of $x(t)$ are independent of the choice of $K$ (one may set $K=1$
for example). 
Equations of the form \eqref{subordination1} are useful simulation tools
\cite{Weron,Weron1}, and were used to investigate multiple-time probability density
functions \cite{Friedrich00,Sokolov,Niemann} and correlation functions
\cite{Friedrich01,Friedrich02}.

\subsection{Two-point probability density functions}

The relation between the solution of the fractional Fokker-Planck equation and
its Brownian counterpart via subordination can be used to derive the
$m$-point probability density functions for a subdiffusion process
\cite{Friedrich00,Sokolov,Niemann}. In particular, the 2-point probability
density function is given by~\cite{Friedrich00}
\begin{eqnarray}
\nonumber
f(x_2,t_2;x_1,t_1)&=&\\
&&\hspace*{-2.8cm}\int_0^{\infty}ds_1\int_0^{\infty}ds_2 
n(s_2,t_2; s_1,t_1)f_M(x_2,s_2;x_1,s_1),
\label{twopoint01}
\end{eqnarray}
where $n(s_2,t_2;s_1,t_1)$ represents the 2-point probability density function
of the inverse L\'{e}vy-stable process $s(t)$, and $f_M(x_2,s_2;x_1,s_1)$ is
the 2-point probability density function of the corresponding Markovian process
$x(s)$ defined in Eq.~\eqref{subordination1}(a). The Laplace transform of $n$ is
given by \cite{Friedrich00}
\begin{eqnarray}
\nonumber
&&n(s_2,\lambda_2; s_1,\lambda_1)=\frac{\exp\left(-s_1\Lambda^\alpha
\frac{K}{K_\alpha}\right)}
{\lambda_1\lambda_2}\\
\nonumber
&&\hspace*{0.4cm}
\times\Big\{\left(\frac{K}{K_\alpha}\right)\delta(s_2-s_1)\left[
\lambda_1^\alpha-\Lambda
^\alpha+\lambda_2^\alpha\right]\\
\nonumber
&&
+\left(\frac{K}{K_\alpha}\right)^2\Theta(s_2-s_1)\lambda_2^\alpha
\left[\Lambda^\alpha-\lambda_2^\alpha\right]e^{-\lambda_2^\alpha
(s_2-s_1)\frac{K}{K_\alpha}}\Big\}\\
\nonumber
&&
\hspace*{2.4cm}
+\frac{\exp\left(-s_2 \Lambda^\alpha 
\frac{K}{K_\alpha}\right)}{\lambda_1\lambda_2}\\
&&
\times\left(\frac{K}{K_\alpha}\right)^2\Theta(s_1-s_2)\lambda_1^\alpha\left[
\Lambda^\alpha-\lambda_1^\alpha
\right]e^{-\lambda_1
^\alpha(s_1-s_2)\frac{K}{K_\alpha}}
\hspace*{0.4cm}
\label{twopoint02}
\end{eqnarray}
where $\Lambda\equiv\lambda_1+\lambda_2$ and $\Theta$ is a step function. Here
and in the following,
the Laplace transform for the variable pairs $t_1\leftrightarrow\lambda_1$ and
$t_2\leftrightarrow\lambda_2$ is denoted by the explicit dependence on the
respective variable.

Knowing that $x(s)$ is a Markov process, the 2-point probability density
function $f_M(x_2,s_2;x_1,s_1)$ of the process $x(s)$ is obtained from the well
known property $f_M(x_2,s_2;x_1,s_1)=P_M(x_2,s_2|x_1,s_1)P_M(x_1,s_1|x_0,0)$,
where
$x_0$ is the initial position of the particle and $P_M(x_1,s_1|x_0,0)$ is
single time probability function for the process $x(s)$ found by solving
Eq.~\ref{subordination1}(a). We decompose this expression in
the form
\begin{eqnarray}
\nonumber
&&f_M(x_2,s_2;x_1,s_1)\\[0.2cm]
\nonumber
&&\hspace*{0.2cm}
=\Theta(s_2-s_1) P_M(x_2,s_2-s_1|x_1,0)P_M(x_1,s_1|x_0,0)\\[0.2cm]
&&\hspace*{0.2cm}
+\Theta(s_1-s_2)P_M(x_1,s_1-s_2|x_2,0)P_M(x_2,s_2|x_0,0).
\hspace*{0.8cm}
\label{twopoint03}
\end{eqnarray}
From Eqs.~\eqref{twopoint01}, \eqref{twopoint02}, and \eqref{twopoint03} we
obtain the subdiffusive probability density function in terms of the Markovian
2-point probability density function $P_M$ in Laplace space via
\begin{eqnarray}
\nonumber
&&f(x_2,\lambda_2;x_1,\lambda_1)\lambda_1\lambda_2 K_\alpha\big/K\\[0.2cm]
\nonumber
&&\hspace*{0.4cm}
=\left[\lambda_1^\alpha-\Lambda^\alpha+\lambda_2^\alpha\right]
\tilde{P}_M(x_1,\frac{K}{K_\alpha}\Lambda^\alpha|x_0,0)\delta(x_2-x_1)\\[0.2cm]
\nonumber
&&
+\frac{K}{K_\alpha}\big\{
\lambda_2^\alpha\left[\Lambda^\alpha-\lambda_2^\alpha\right]
\tilde{P}_M(x_2,\frac{K}{K_\alpha}\lambda_2^\alpha|x_1,0)\tilde{P}_M(x_1,
\frac{K}{K_\alpha}\Lambda^\alpha|x_0,0)
\\[0.2cm]
&&
+
\lambda_1^\alpha\left[\Lambda^\alpha-\lambda_1^\alpha\right]
\tilde{P}_M(x_1,\frac{K}{K_\alpha}\lambda_1^\alpha|x_2,0)\tilde{P}_M(x_2,
\frac{K}{K_\alpha}\Lambda^\alpha|x_0,0)\big\}.
\hspace*{0.8cm}
\label{twopoint04}
\end{eqnarray}
Here $\tilde{P}_M(x,\frac{K}{K_\alpha}\lambda^\alpha|x_0,0)$ denotes the Laplace
transform 
($\frac{K}{K_\alpha}\lambda^\alpha\leftrightarrow s$) of $P_M(x,s|x_0,0)$,
and similarly for $\Lambda^{\alpha}$.
Eq.~\eqref{twopoint04} is the final result connecting the 2-point probability
density function $f$ in Laplace space to its Markovian counterpart.

\section{Results}

\subsection{Position autocorrelation}

\begin{figure}
\includegraphics[width=8.2cm]{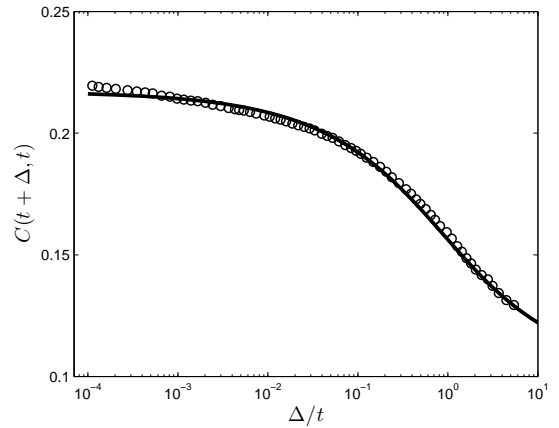}
\caption{Position correlation correlation function $C(t_2,t_1)=\langle x(t_2)
x(t_1)$ from Eq.~\eqref{corr_sol02} (solid line) as function of the scaled
time $\Delta/t$ for $\alpha=0.525$, where $\Delta=|t_2-t_1|$. The values of
the Boltzmann moments are $\langle x^2 \rangle_B = 0.217$ and $\langle x
\rangle_B^2=0.102$. We also show data ($\circ$ symbols) for the magnetization
$M(t+\Delta,t)$ from the thermoremanent magnetization experiment reported
in Ref.~\cite{Orbach} from
measurement in a $\text{Cu}_{0.94}\text{Mn}_{0.06}$ sample, showing aging
behavior. The shown data correspond to the longest measured waiting time,
$t=10000$ sec.
\label{fig2}}
\end{figure}

 Eq.~\eqref{twopoint04} allows one to calculate general two-point correlation
functions for subdiffusive systems governed by the fractional Fokker-Planck
equation \eqref{ffpe01}. In particular, the position-position
correlation function is
\begin{equation}
\langle x(\lambda_2)x(\lambda_1)\rangle=
\int_{-\infty}^{\infty}\int_{-\infty}^{\infty}x_1x_2f(x_2,\lambda_2;x_1,
\lambda_1)dx_1dx_2 
\end{equation}
in Laplace space. Laplace inversion (see Ref.~\cite{unpublished} for
details) delivers the final result for the 2-time correlation function
\begin{equation}
\langle x(t_2)x(t_1)\rangle\sim\Big(\langle x^2\rangle_B-\langle x\rangle_B^2
\Big)\frac{B(t_1/t_2,\alpha,1-\alpha)}{\Gamma(\alpha)\Gamma(1-\alpha)}
+\langle x\rangle_B^2,
\label{corr_sol02}
\end{equation}
valid for $t_2\geq t_1\gg(1/K_{\alpha}\mu_1)^{1/\alpha}$, $\mu_1$ being
the smallest non-zero eigenvalue of the Fokker-Planck operator
$\mathbb{L}_{\mathrm{FP}}$~\cite{Risken}, and
\begin{equation}
B(z,a,b)=\int_0^z y^{a-1}(1-y)^{b-1}\,dy
\end{equation}
is the incomplete Beta function \cite{Abramowitz}. The symbol $\langle\cdot
\rangle_B$ denotes an average over the Boltzmann distribution 
\begin{equation}
\langle x^n\rangle_B=\int_{-\infty}^{\infty}x^n e^{-U(x)/k_BT}dx\Big/\int_{-
\infty}^{\infty}e^{-U(x)/k_BT}dx.
\end{equation}
Fig.~\ref{fig2} shows the sigmoidal behavior of the position autocorrelation
function. It is important to
note that Eq.~(\ref{corr_sol02}) implies that the process is irreversible
since in the limit $\Delta\to\infty$ $B(t\big/t+\Delta,\alpha,1-\alpha)\to
0$ and $\langle x(t_2)x(t_1)\rangle\to\langle x\rangle_B^2$.

The result \eqref{corr_sol02} has some remarkable properties. Thus the external
potential $U(x)$ enters only via the prefactor, and only through the first two
moments of the corresponding Boltzmann distribution. The long time behavior of
$\langle x(t_2)x(t_1)\rangle$ is universal and depends only on the ratio $t_2/
t_1$. 

We note that Beta function behavior for a correlation function was found
previously for a simple two-state renewal model
with power-law sojourn times on both
states~\cite{Glasses,Dahan,margo,margo1,Godreche,Dean}. 
While our process is clearly not a two state process, the
universal behavior of expression
\eqref{corr_sol02} is due to the separation between the physical process in
space and the associated temporal process. Such a separation is exactly the
idea behind the subordination of time, Eq.~\eqref{subordination1}(b). The
temporal process $t(s)$ yields the time evolution governed by the waiting
times between successive jumps. Due to the assumption of annealed disorder it
is independent of the current particle position. It converges as a function
of the number of jumps ($\propto s$) due to the generalized central limit
theorem, corresponding to the long-time limit $s\to\infty$ in
Eq.~\eqref{subordination1}(b). The limiting behavior of $t(s)$ is therefore a
L{\'e}vy stable law which is underlying the subdiffusion dynamics. Conversely,
the spatial process explores the external potential and is not affected by the
disorder if observed as a function of the internal time $s$. In fact, as
function of $s$ the process corresponds to normal diffusion in an external
field, and so the process converges to Boltzmann statistics characterised by
the binding properties of the external potential. We note that the result
\eqref{corr_sol02} for the correlation function mirrors the convergence of
both temporal and spatial processes, and is independent of the microscopic
properties of the model (e.g. the shape of $U(x)$). To obtain the behavior when
one of the processes has
not converged one needs to use the full correlation function with a non-trivial
time dependence \cite{unpublished}. The latter depends on all eigen-values of 
$\mathbb{L}_{\mathrm{FP}}$ and is cumbersome.

\subsection{Properties of the two-time position correlation}

The correlation function \eqref{corr_sol02} displays a number of noteworthy
features:

\subsubsection{(i) Aging behavior}

The correlation function $C(t+\Delta,t)=\langle x(t+\Delta)x(t)\rangle$
exhibits
aging since its time dependence is of the form $\Delta/t$. Aging behavior was
indeed observed in many complex systems \cite{Glasses,Weitz,Orbach}, for
instance, in thermoremanent magnetization experiments \cite{Ocio,Orbach}, in
which the measured relaxation of the magnetization $M(t+\Delta,t)$ is
proportional to the spin correlation function, according to generalized
fluctuation-dissipation theorems \cite{Coglindolo,Dean,Glasses}.
Accepting
our stochastic theory as an approximation for the spin system behavior  
we used Eq.~\eqref{corr_sol02} to fit the thermoremanent magnetization data
from Ref.~\cite{Orbach}. The result of the fit is presented in Fig.~\ref{fig2}.
We observe good agreement between the data and our Beta function results over
the entire time window, with a slight discrepancy at short times. We note that
the use of a non-zero value for the fitting parameter $\langle x\rangle_B^2$
in Eq.~\eqref{corr_sol02} for the zero external field relaxation of the
magnetization is consistent with observed asymmetric magnetic fluctuations in
thermoremanent magnetization experiments~\cite{Kenning} as opposed to the
naively expected zero average behavior. 
Fitting with the Beta function for the measured correlation function does
not necessarily yield insight into the physics of the system, but
classification of aging with particular fitting functions might be profitable
step (as the well known functions, such as Cole-Cole plots, are useful in the
classification of dielectric relaxation).

\subsubsection{(ii) Time-averaged position}

We quantify ergodicity, or the departure from ergodicity, of the system by
measuring fluctuations of the time averaged position
\begin{equation}
\overline{x(t)}=\frac{1}{t}\int_0^t x(t')\,dt'.
\end{equation}
Combining Eqs.~\eqref{ergod5} and \eqref{corr_sol02} we see that
\begin{equation}
\lim_{t\to\infty}\left<\overline{x(t)}^2\right>=(1-\alpha)(\langle x^2
\rangle_B -\langle x\rangle_B^2)+\langle x\rangle_B^2,
\label{web01}
\end{equation}
 where we used the relation
\begin{equation}
\int_0^1 B(z,\alpha,1-\alpha)dz=(1-\alpha)\Gamma(\alpha)\Gamma(1-\alpha).
\end{equation}
This result was previously obtained for a CTRW process \cite{Robenshtok}.
Clearly $\left<\overline{x(t)}^2\right>\neq\langle x\rangle_B^2$ when
$\alpha<1$, thus we observe weakly non-ergodic behavior. In the limit
$\alpha=1$ ergodicity is restored.

\subsubsection{(iii) Edwards-Anderson parameter}

The Edwards-Anderson parameter was previously used to quantify the degree of
weak ergodicity breaking in the context of spin-glasses~\cite{Glasses}. It is
defined as 
\begin{equation}
q_{EA}=\lim_{\Delta\to\infty}\lim_{t \to \infty}C(t+\Delta,t).
\end{equation}
In our current framework, for the case of a symmetric potential the
Edwards-Anderson parameter becomes
\begin{equation}
q_{EA}=\left\{\begin{array}{ll} \langle x^2\rangle_B, & \alpha<1\\
0, & \alpha\ge1 
\end{array}
\right.,
\label{qparametr}
\end{equation}
 reflecting irreversibility of our process which is still
non-ergodic. Conversely, interchanging the limits we find
that
\begin{equation}
\lim_{t\to\infty}\lim_{\Delta\to\infty}C(t+\Delta,t)=0.
\end{equation}
reflecting the aging character of the system. Note that Eq.~\eqref{qparametr}
indicates that $q_{EA}$ is determined by the Boltzmann distribution and is
independent of $\alpha$ in the nonergodic phase.


\subsubsection{(iv) Time-averaged mean squared displacement}

From Eq.~\eqref{corr_sol02} we also obtain the behavior of the ensemble
average of the time averaged mean squared displacement \cite{Klafter,He}.
Namely, from a time series $x(t)$ recorded in single particle tracking
experiments one can define the time averaged mean squared displacement
\begin{equation}
\overline{\delta^2(\Delta)}=\frac{1}{T-\Delta}\int_0^{T-\Delta}\Big[x(t
+\Delta)-x(t)\Big]^2dt,
\end{equation}
where $T$ is the overall measurement time. At finite measurement time $T$
even in the Brownian limit the quantity $\overline{\delta^2(\Delta)}$ is
a random quantity depending on the particular trajectory. Performing an
additional ensemble average, for a Brownian system the role of the lag time
$\Delta$ in the long measurement time limit is completely interchangeable
with the regular $t$-dependence in the corresponding ensemble averaged mean
squared displacement, for example when $U(x)=0$
$\lim_{T\to\infty}\overline{\delta^2(\Delta,T)}=2K\Delta$. In presence of a
confining potential one would naively
expect the mean squared displacement to saturate, as observed for a Brownian
system. However, evaluating the ensemble average of $\overline{\delta^2(
\Delta)}$ 
\begin{eqnarray}
\nonumber
\left<\overline{\delta^2(\Delta)}\right>=\frac{1}{T-\Delta}\int_0^{T-\Delta}
\left<\Big[x(t+\Delta)-x(t)\right]^2\Big>dt\\[0.2cm]
=\frac{{\displaystyle\int}_0^{T-\Delta}\Big[\langle x^2(t+\Delta)\rangle+\langle
x^2(t)\rangle-2C(t+\Delta,t)\Big]dt}{T-\Delta}.
\label{tamsd}
\end{eqnarray}
we find the a priori surprising result :
For regular diffusion in a binding potential one obtains a saturation for
long times, as for anomalous motion in the case of the ensemble average
\cite{MetzlerKlafter,report}. In contrast, for the time averaged mean squared
displacement in our anomalous system from Eq.~\eqref{tamsd} we find
from the Beta function expansion
\begin{equation}
\frac{B(t/(t+\Delta),\alpha,1-\alpha)}{\Gamma(\alpha)\Gamma(1-\alpha)}\sim
1-\frac{\sin(\alpha\pi)}{(1-\alpha)\pi}\left(\frac{\Delta}{t}\right)^{1-\alpha}
\end{equation}
the behavior
\begin{equation}
\left<\overline{\delta^2(\Delta)}\right>\sim\Big(\langle x^2\rangle_B-
\langle x\rangle_B^2\Big)\frac{2\sin(\alpha\pi)}{(1-\alpha)\alpha\pi}
\left(\frac{\Delta}{T}\right)^{1-\alpha},
\label{msd01}
\end{equation}
valid in the limit $(\Delta/T)\ll1$ and for
$\Delta\gg(1/K_\alpha\mu_1)^{1/\alpha}$
\cite{unpublished}. Instead of the naively expected saturation, the time
averaged mean squared displacement grows as a power with exponent $(1-\alpha)$.
Only when the lag time approaches the measurement time $T$ this power-law
growth stops, and the function dips to the ensemble averaged value. We note
that the $\Delta^{1-\alpha}$ scaling was recently reported for the case
of a particle in a box \cite{igor}.

\subsubsection{(v) Numerical analysis of position autocorrelation.}

\begin{figure}
\includegraphics[width=8cm]{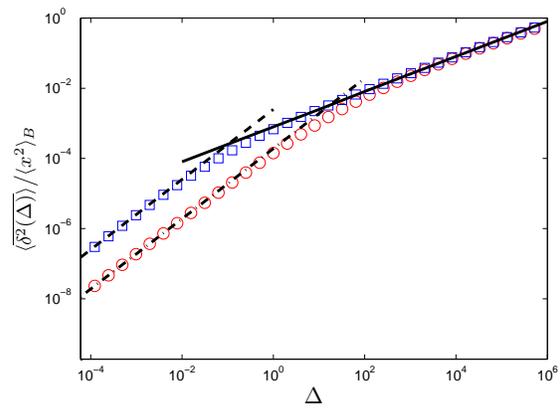}
\caption{Simulated behavior of the time-averaged mean squared displacement
$\left<\overline{\delta^2(\Delta)}\right>$ for a continuous time random
walk process with waiting time distribution $\psi(t)\sim\tau^{\alpha}/t^{
1+\alpha}$, in an harmonic binding potential $U(x)=x^2$ ($\square$), and in
a box with reflecting boundary conditions and size $2$ ($\circ$). The
anomalous diffusion exponent is $\alpha = 1/2$, and the measurement time
was chosen as $T=10^7$ (a.u.). We also chose $k_BT_{mp}=0.1$, and
$K_{1/2}=0.0892$.
Without fit, the lines show the analytic results for the transition from the
initial linear lag time dependence $\simeq\Delta^1$, Eq.~\eqref{freepart}
(--- and $-\cdot-$), to the long lag time behavior $\simeq\Delta^{1-\alpha}$,
Eq.~\eqref{msd01} (---). In both cases $\langle x\rangle_B=0$. At long lag
times $\left<\overline{\delta^2(\Delta)}\right>/\langle x^2\rangle_B$ exhibits
universal behavior, independent of the external field.
\label{fig1}}
\end{figure}

Fig.~\ref{fig1} shows, over a large time span, the time-averaged mean squared
displacement $\left<\overline{\delta^2(\Delta)}\right>$ of a subdiffusing
particle in (i) an harmonic potential, and (ii) in a box with reflecting
boundaries. The initial particle position was chosen to be at the bottom of
the potential and in the center of the box, respectively.
At short lag times $\Delta$ we observe the linear scaling
\begin{equation}
\left<\overline{\delta^2(\Delta)}\right>
\sim\frac{2K_\alpha\Delta}{\Gamma(1+\alpha)T^{1-\alpha}}
\label{freepart} 
\end{equation}
with the lag time $\Delta$. In this result only the dependence on the overall
measurement time $T$ bears witness to the fact that the underlying stochastic
process is subdiffusive. Seemingly paradox, the lag time $\Delta$ enters
linearly, in contrast to the associated ensemble averaged mean squared
displacement $\langle x^2(t)\rangle\sim 2K_{\alpha}t^{\alpha}/\Gamma(1+\alpha)$.
However,
this is the result of the weak ergodicity breaking of the process, as shown
in Refs.~\cite{Klafter,He}. The free particle behavior at short $\Delta$
is an expected result, which can be obtained from scaling arguments or
explicitly from the full correlation function \cite{unpublished}: at
sufficiently short times the particle does not yet feel the confinement due
to the reflecting boundaries, or it does not yet experience the restoring
force of the potential, respectively. The $\left<\overline{\delta^2(\Delta)}
\right>\simeq\Delta$ regime holds for scales of the lag time $\Delta$
that fulfill $K_\alpha\Delta^\alpha \ll L^2$ in the example of the box, where
$L$ is
the size of the box. For a general confining potential, the turnover time is
non-universal and is dependent on all non-zero eigenvalues $\mu_n$ of the
Fokker-Planck operator $\mathbb{L}_{\mathrm{FP}}$ \cite{Risken}. Thus, at
times $\Delta\gg(1/K_\alpha\mu_1)^{1/\alpha}$ a transition occurs to the $\left<
\overline{\delta^2(\Delta)}\right>\sim\Delta^{1-\alpha}$ regime,
Eq.~\eqref{msd01}. We stress again that, in contrast to normal diffusion, no
saturation is found at long lag times, and $\left<\overline{\delta^2(\Delta)}
\right>$ continues to grow for any $\Delta<T$. Only as $\Delta$ approaches
to the measurement time $T$, we obtain the convergence
$\left<\overline{\delta^2(\Delta)}\right>\to\langle x^2\rangle+\langle
x^2(0)\rangle$.

\begin{figure}
\includegraphics[width=8cm]{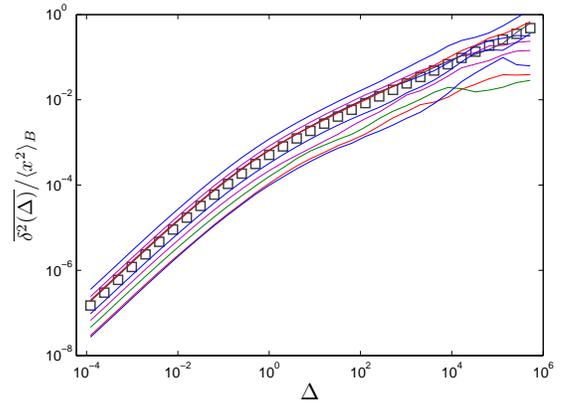}
\caption{Single trajectories for the motion in the harmonic binding potential
$U(x)=x^2$, for the same parameters as in Fig.~\ref{fig1} (thin lines). The
symbols ($\square$) represent the simulations result for $\left<\overline{
\delta^2(\Delta)}\right>$. The scatter between individual trajectories is
distinct and resembles qualitatively typical experimental results
\cite{Golding,b1,b2,b3,b4,b5}.
\label{fig3}}
\end{figure}

In Fig.~\ref{fig3} we show the simulations result for a number of individual
trajectories in an harmonic binding potential, displaying significant scatter.
This scatter between individual trajectories is, again, a result of the weak
ergodicity breaking of the underlying stochastic process, and resembles
qualitatively the behavior observed in single particle tracking experiments
\cite{Golding,b1,b2,b3,b4,b5}. The amplitude of the
scatter can be quantified in terms of the dimensionless random variable
$\xi\equiv\overline{\delta^2(\Delta)}/\left<\overline{\delta^2(\Delta)}
\right>$, the relative scatter of the time average with respect to its
ensemble mean. Using arguments similar to~\cite{He} we can
show~\cite{unpublished} that the distribution of $\xi$ is given in terms of a
one-sided
L{\'e}vy stable law in the form \cite{He}
\begin{equation}
\lim_{T\to\infty}\phi_{\alpha}(\xi)=\frac{\Gamma(1+\alpha)^{1/\alpha}}{
\alpha\xi^{1+1/\alpha}}l_{\alpha}\left(\frac{\Gamma(1+\alpha)^{1/\alpha}}{
\xi^{1/\alpha}}\right),
\label{scatter01}
\end{equation}
where the stable density $l_{\alpha}(t)$ is defined in terms of its Laplace
image $\exp(-u^{\alpha})$. Note that the random variable $\xi$ is in the
denominator of $l_{\alpha}$, and therefore the associated width is finite.
For instance, for the case $\alpha=1/2$ used in Figs.~\ref{fig1} and
\ref{fig3} we find the Gaussian law $\lim_{T\to\infty}\phi_{1/2}(\xi)=
(2/\pi)\exp\left(-\xi^2/2\right)$.
Finally, in the Brownian limit $\alpha=1$, the distribution converges to the
sharp behavior $\phi_1(\xi)=\delta(\xi-1)$, restoring ergodicity in the sense
that no more scatter occurs. The distribution of $\xi$, given by
Eq.~(\ref{scatter01}), is the same for both unbounded and bounded anomalous
diffusion. This is simply due to the mentioned fact that
temporal and spatial process are uncoupled.

\section{Discussion}

Correlation functions are a standard tool to experimentally probe the temporal
evolution of a system. They provide information on how the present value of a
physical observable influences its value in the future, and are therefore
important indicators to the specific process that governs the system's
dynamics. 
The significance of the correlation function behavior for the
fundamental concepts in physics is revealed through Khinchin's theorem which
provides a condition for a stationary process to be ergodic in terms of the
corresponding correlation function. Herein we derived a generalization of
Khinchin's theorem for a class of non-stationary aging process. We provide not
only a generalized condition for ergodicity for such processes in terms of the
corresponding aging correlation function but also quantify the
deviations from ergodicity.
For the broad class of non-stationary processes described by Eq.~\eqref{ffpe01}
we derived
analytically the time dependence of two-point
correlation functions for subdiffusing particles under situations of
confinement. In particular we revealed a universal behavior for the two-time
position correlation function involving the incomplete Beta function. All
features of the confining potential enter the correlation function solely
through the prefactor in terms of the first and second moments of the
associated Boltzmann distribution. Of course,
the expression for the correlation function in Eq.~\eqref{corr_sol02}
is not restricted to a position correlation function, but can be used to
describe a very general class of quantities, e.g., a potential energy
correlation function. The generality of our results is a direct consequence of
the
convergence of
the Markovian process $x(s)$ in the jump space, and the ubiquitous role
of L\'{e}vy statistics due to the generalized central limit theorem.


We note that the behavior found for the time-averaged mean
squared displacement in binding potential (or confined geometry) recovered
herein could be
verified in a single particle tracking experiment specifically for
micro-movement inside a yeast cell, in which the particle
was tracked indirectly in an optical tweezers setup, that eventually
exerts a Hookean restoring force on the tracer particle \cite{lene}. 
Similarly anomalous diffusion of macro molecules in a cell is always confined by
the cell boundary, indicating that our theory of confined anomalous diffusion
may have wide applications.

\begin{acknowledgments}
This work was supported by the Israel Science Foundation and the Deutsche
Forschungsgemeinschaft. We thank G.G. Kenning for permitting us to use the
experimental data published in Ref.~\cite{Orbach}.
\end{acknowledgments}

\end{article}

\end{document}